\documentclass[superscriptaddress,twocolumn,showpacs,preprintnumbers,amsmath,amssymb,prl,floatfix]{revtex4}

\usepackage{graphicx}
\usepackage{bbm}
\usepackage{subfigure}

\def\>{\rangle}
\def\<{\langle}

\def\Tr{{\text{Tr}}}
\def\env{{\text{env}}}
\def\one{\openone}
\def\d{\text{d}}
\def\e{{\textrm{e}}}
\def\q{{\textrm{q}}}

\begin{document}
\title{A Bell pair in a generic random matrix environment}

\author{Carlos Pineda}
\email{carlospgmat03@gmail.com}
\affiliation{Instituto de F\'{\i}sica, UNAM, M\'exico}
\affiliation{Centro de Ciencias F\'{\i}sicas, UNAM, M\'exico}
\author{Thomas H. Seligman}
\affiliation{Centro de Ciencias F\'{\i}sicas, UNAM, M\'exico}
\affiliation{Centro Internacional de Ciencias, Cuernavaca, M\'exico}
\date{\today}
\begin{abstract}
  Two non-interacting qubits are coupled to an environment. Both coupling and
  environment are represented by random matrix ensembles. The initial 
  state of the pair is a Bell state, though we also consider arbitrary pure
  states. Decoherence of the pair is evaluated analytically in terms of
  purity; Monte Carlo calculations confirm these results and also yield the
  concurrence of the pair.  Entanglement within the pair
  accelerates decoherence.  Numerics display the relation between concurrence
  and purity known for Werner states, allowing us to give a formula
  for concurrence decay. 
\end{abstract}


\maketitle

The evolution of entanglement within a pair of qubits or spin $1/2$ particles
under the influence of an environment is paradigmatic for the stability of
teleportation \cite{specialteleport}, and indeed for any quantum information
process \cite{NC00a}.  Concurrence provides a measure for the degree of
entanglement within such a pair \cite{firstconcurrence}. The main purpose of
the present paper is to establish the generic behavior of the decay of
entanglement, {\it i.e.}  concurrence, of such a pair of qubits.  We propose a
random matrix model for the unitary evolution of a pair of qubits interacting
with one or two environments, but not among themselves.  The environment(s) as
well as the couplings will be described by one of the classical ensembles of
random matrices \cite{cartanRMT}. Research in "quantum chaos" has revealed,
that such ensembles describe a chaotic environment well
\cite{guhr98random,muller:046207} and relations to the Caldeira-Legget model have 
been established \cite{caoslutz}.
In the present article we shall concentrate on the Gaussian unitary ensemble
(GUE), which describes time-reversal breaking dynamics, mainly because it
provides the simplest analytics.  The model we use is based on one developed
in general for the evolution of decoherence \cite{1464-4266-4-4-325} and
applied to fidelity decay \cite{1367-2630-6-1-020}; the latter was
successfully tested by experiment
\cite{expRudishort,njpRudishort,gorin-weaver}.  We shall also need information
about the evolution of the entanglement of the pair with the environment {\it
  i.e.}  of its decoherence.  This we measure in terms of purity \cite{Zur91}
rather then von Neumann entropy, because the analytic structure of purity
allows an analytic treatment in terms of a Born expansion. Using this
expression we see that purity of an entangled state decays faster than purity
of a product state, but we shall be able to go one step further. Numerically
we show that the relation of purity to concurrence demonstrated for a specific
dynamical model \cite{pineda:012305} is universal, in the sense that it holds
for the random matrix model.  This relation coincides with the one for a
Werner state and thus is analytically known, allowing us to give a closed,
though heuristic, expression for concurrence decay.  Both quantities and thus
their relation are acessable by quantum tomography in experiments with trapped
ions or atoms, whereinteraction with a controled environment is feasable
\cite{blattprivate}.

Concurrence of a density matrix $\rho$ representing the state of a pair of
qubits, is defined as
\begin{equation}
\label{eq:concurrence}
C(\rho)=\max \{0,\lambda_1-\lambda_2-\lambda_3-\lambda_4 \}
\end{equation}
where $\lambda_i$ are the eigenvalues of the matrix $\sqrt{\rho (\sigma_y
  \otimes \sigma_y) \rho^* (\sigma_y \otimes \sigma_y)}$ in non-increasing
order; $(^*)$ denotes complex conjugation in the computational basis and
$\sigma_y$ is a Pauli matrix.  Purity is defined as
\begin{equation}
\label{eq:purity}
P(\rho)=\Tr \rho^2.
\end{equation}

We study dynamics on a Hilbert space with the structure 
${\mathcal H}\ =\ {\mathcal H}^\q_1 \otimes {\mathcal H}^\e_1 \otimes
{\mathcal H}^\q_2 \otimes {\mathcal H}^\e_2$, where ${\mathcal H}^\q_i $
indicates (two dimensional) qubit spaces, while ${\mathcal H}^\e_i$ will
indicate $N$ dimensional environments.

We consider unitary dynamics on the entire space and obtain the
non-unitary dynamics for the qubits by partial tracing over the environment(s).
As we wish to consider the effect of the environment on the pair of qubits, we
cannot allow any interaction within the pair, but we consider
interactions with the environments, which may be fused to a single one. 
For convenience we also neglect any possible evolution for each qubit
individually, which is not induced by the coupling to the environment. The
latter is non-essential to our argument, but simplifies the analytic
treatment.  We thus use the Hamiltonian
\begin{equation}
\label{eq:hamiltonianboth}
H= H_1^\e + H_2^\e +\lambda_1 V_1^{\e,\q}+\lambda_2 V_2^{\e,\q}.
\end{equation}
The first two terms correspond to dynamics of the environments. The third and
fourth terms represent the coupling of each of the qubits to the corresponding
environment.  To obtain further simplification, we consider one of the
qubits as a \textit{spectator}, \textit{i.e.} we assume that it has no coupling
to an environment ($\lambda_2=0$).  The corresponding environment becomes
irrelevant and we obtain the simplified Hamiltonian
\begin{equation}
\label{eq:hamiltonianwitness}
H_\lambda= H^\e_1 +\lambda V^{\e,\q}_1.
\end{equation}
Note that we do consider entanglement with the spectator.  This yields the
\textit{simplest} Hamiltonian for which we can analyze the effect of an
environment on a Bell pair.  The environment Hamiltonian $H^\e_1$ will be
chosen from a classical ensemble~\cite{cartanRMT} of $N \times N$ matrices and
the coupling, $V^{\e,\q}_1$, from one of $2N \times 2N$ matrices.  As usual,
the GUE ensemble, which represents time-reversal invariance breaking dynamics,
is easier to handle analytically than the Gaussian orthogonal one. Here we
focus on the former, while treating the latter in a follow up paper. Evolution
of both purity and concurrence of the pair of qubits can readily be simulated
in a Monte Carlo calculation and due to the simple structure of purity, it is
possible to compute analytically this quantity in linear response
approximation.  An exact calculation requires four-point functions, which
despite of the power of super-symmetric techniques~\cite{guhr98random} are
still not readily available.

To calculate the value of purity, we use the following averages and
approximations. First we expand the evolution operator as a Born series; hence
we require small $\lambda$ and/or short times. We average both $V_{1}^{\e,\q}$
(which will be called $V$ from now on) and $H^\e_1$ over the appropriate GUE
ensemble.  Finally we average the initial state and obtain
Eq.~\eqref{eq:both}.  This is the very same scheme followed
in~\cite{1367-2630-6-1-020} for fidelity decay, though details are more
complicated~\cite{1464-4266-4-4-325} due to the partial traces.

We define the evolution operator $U_\lambda=\exp[-\imath H_\lambda t]$, such
that the density matrix in Eqs.~\eqref{eq:concurrence} and \eqref{eq:purity}
is $\rho(t)=\Tr_\env U_\lambda |\psi(0)\>\<\psi(0)|U_\lambda^\dagger$, where
$|\psi(0)\>$ is the initial state of the system.  Since $U_0$ is a local
operation in the environment it will not affect the value of $\rho$. Thus we
can equally evolve with $U_0^\dagger U_\lambda$ instead of $U_\lambda$ alone.
It is convenient to use $U_0^\dagger U_\lambda$ since for small $\lambda$ this
operator will remain in some sense near to unity for longer times. We write
the Born series to second order as
\begin{equation}   \label{eq:born}
U_0^\dagger \, U_{\lambda} \approx \one-\imath \lambda I (t) - \lambda^2 J(t);
\end{equation}
\begin{equation} 
I(t)= 
 \int_0^t \d \tau \tilde V(\tau) ; \, J(t)= \int_0^t \d \tau \int_0^\tau \d \tau' \tilde V(\tau) \tilde V(\tau').
\end{equation}
Here $\tilde{V}(t)$ is the coupling operator in the interaction picture:
$\tilde{V}(t)=U_0^\dagger V U_0$.
Writing $|\psi(0)\>=\sum_{\mu=1}^{4  N}x_\mu |\mu\>$,
and using Eq.~\eqref{eq:born}, purity reads as:
\begin{equation}   \label{eq:PurityAfterBorn}
P(t) \approx 1- \lambda ^2 (\text{Re} A_J -A_1-A_2+\text{Re} A_3) ;
\end{equation}
\begin{subequations}
\begin{align}
A_J &=4 x_\mu x_{i'jk}^* x_{i'j'k^\prime} x_{ij'k^\prime}^* J_{ijk,\mu}(t),   \\
A_1&=2  x_\mu x_\nu^* x_{i'j'k^\prime} x_{ij'k^\prime}^* I_{ijk,\mu}(t)I_{i'jk,\nu}^*(t),\\
A_2&=2  x_\mu x_{i'jk}^* x_{i'j'k^\prime} x_\nu^* I_{ijk,\mu}(t) I_{ij'k',\nu}^*(t),  \\
A_3&=2  x_\mu x_{i'jk}^* x_\nu x_{ij'k^\prime}^* I_{ijk,\mu}(t) I_{i'j'k',\nu}(t)
\end{align}
\end{subequations}
(summation over repeated indices is assumed).  Indices run as follows: Greek
ones over the whole Hilbert space, the $i$'s over the environment, $j$'s over
the first qubit and $k$'s over the spectator qubit. 
Note that we use the natural notation for the indices of vectors in a space
which is a tensor product of several spaces.

We now average the perturbation $V$ over the GUE using 
$\left\langle V_{m,n} \right\rangle=0$ and $\left\langle V_{m,m^\prime}
  V_{n,n^\prime} \right\rangle= \delta_{m,n^\prime}\delta_{m',n}$. 
Due to the unitary invariance of the GUE we choose the basis that diagonalizes 
$H^\e_1$ yielding eigenvalues $E_i$. Then
\begin{multline}
\label{eq:J1}
\< J(t) \>_{ijk,i'j'k^\prime}=2 \delta_{ijk,i'j'k^\prime}\\
\times \int_0^t \d \tau \int_0^\tau \d \tau'\sum_{i''} e^{\imath (\tau'-
\tau)(E_{i''}-E_i)}.
\end{multline}
The matrix elements of the tensors $I\otimes I$ and $I\otimes
I^*$, averaged, yield
\begin{align*}
  \label{eq:II}
  \<&I_{ijk,lmn}(t_1)I_{i'j'k',l'm'n^\prime}(t_2) \>\\
    &\quad= -\<I_{ijk,lmn}(t_1)I_{l'm'k',i'j'n^\prime}^*(-t_2) \>\\
    &\quad=\delta_{ijk,l'm'n} \delta_{i'j'k',lmn'}
      \int_0^{t_1} \d\tau \int_0^{t_2} \d \tau' e^{\imath (E_i-E_l)(\tau-\tau')}.
\end{align*}
Next we average $H^\e_1$ over the GUE using that $\langle
\sum_{i,i'}e^{\imath (E_i-E_{i'}) t} \rangle =N
[1+\delta(t/\tau_H)-b_2(t/\tau_H)]$, where $ b_2(t/\tau_H)$ is the form factor
of the GUE~\cite{guhr98random} and $\tau_H$ the Heisenberg time, set to $2\pi$
throughout this letter. 

\begin{figure}
     \includegraphics{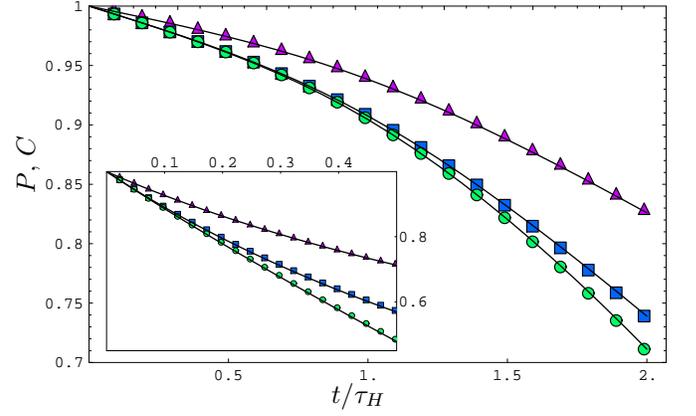}
     \caption{We show the evolution of purity for separable states
       (purple triangles), Bell states (green circles) and concurrence for
       Bell states (blue squares) in the crossover regime.  The lines show theoretical predictions
       given by Eqs.\eqref{eq:purityGUE} and \eqref{eq:goodexponwernerCtime}.
       The environment has dimension $2^{10}$ and the perturbation strength is
       $\lambda=0.025$.  In the inset we  observe the Fermi golden rule regime
       for a larger perturbation $\lambda=0.1$.}
  \label{fig:TransitionTime}
\end{figure}
The initial state is a product of pure states for the qubit pair and the
environment. For the latter we use a random initial state $\sum_i x_i |i\>$,
constructed in the large $N$ limit, using complex random numbers $x_i$
distributed according to a Gaussian centered around zero with width
$1/\sqrt{N}$.  For the pair of qubits we choose a completely general pure
state. Since we still have the freedom to perform an arbitrary unitary local
operation on each qubit, we pick a basis in which the initial state for the
two qubits is 
\begin{equation}
  \label{eq:simpleintialstate}
  |\varphi_\alpha\>=\cos \alpha|00\>+\sin \alpha|11\>, \,\alpha\in[0,\pi/4]. 
\end{equation}
The degree of entanglement is characterized by $\alpha$; in fact
$C(|\varphi_\alpha \>\< \varphi_\alpha|)=\sin2\alpha$. Hence our initial state
can be written as $|\psi(0)\>=\sum_i x_i |i\>|\varphi_\alpha\>$.
Neglecting higher order terms in $1/N$, we obtain 
$\< \<A_J \>\>=2 f(t)$ and $\< \< A_2 \>\>=g_\alpha f(t)$ with
\begin{equation}
\label{eq:resfinf}
f(t)=\begin{cases} 
             2t\tau_h + \frac{2t^3}{3\tau_h} &
                   \text{if $0\leq t<\tau_H$}\\[.2cm]
             2t^2+\frac{2\tau_h^2}{3}&\text{if $t\geq\tau_H $},
           \end{cases}
\end{equation}
and $g_\alpha=\cos^4\alpha + \sin^4\alpha$.
To leading order $\< \< A_1 \>\>=\< \< A_3 \>\>=0$. We obtain
\begin{equation}\label{eq:both}
P_\textrm{LR}(t) = 1- \lambda^2 (2-g_\alpha) f(t).
\end{equation}
From this result we see directly that purity decay will be faster the more
entangled the initial state was.  The validity of this approximation is
limited to large values of purity, \textit{i.e.}  short times or weak
coupling. This is valuable for applications to quantum information, but
we are interested in the dynamical picture as a whole and thus would like to
obtain an expression valid for a wide range of physical situations.  As a way
to achieve this for fidelity decay, exponentiation of the leading term of the
linear response formula was proposed~\cite{purityfidelity}. A similar
approximation is taken here, to obtain $P_\textrm{ELR}(t)$.  In order to
calculate the appropriate formula, we must satisfy $P_\textrm{ELR}(t) \approx
P_\textrm{LR}(t)$ for small $t$, and consider correct asymptotics.  These will
be estimated as the purity after applying a totally depolarizing channel on
one qubit to the 2 qubit state Eq.~\eqref{eq:simpleintialstate}.  The expected
asymptotic value is $g_\alpha/2$, and the final expression is
\begin{equation}
  \label{eq:purityGUE}
  P_\textrm{ELR}(t) = \frac{g_\alpha}{2}+\left(1- \frac{g_\alpha}{2}\right)
         e^{\frac{P_\textrm{LR}(t)-1}{1-g_\alpha/2}}.
\end{equation}
This result is in excellent agreement with numerics as shown in
Fig.~\ref{fig:TransitionTime}, and displays the transition from
exponential to Gaussian decay as Heisenberg time is reached.  

\begin{figure}
     \includegraphics{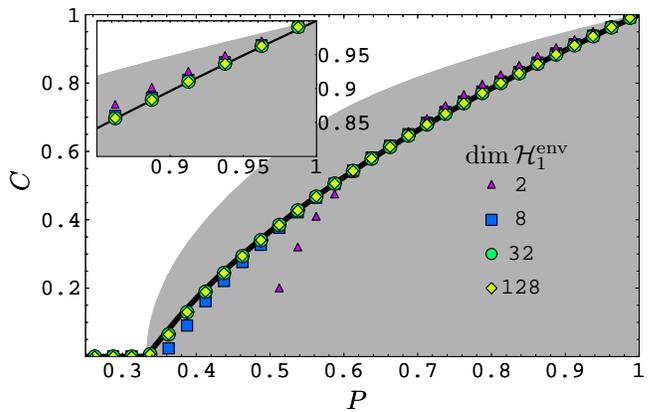}
     \caption{The relation between $C$ and $P$ for a fixed
       coupling ($\lambda=0.3$) and variable size of the 
       environment.     We     average     over     10 realizations  of the Hamiltonian
       \eqref{eq:hamiltonianwitness}  and  15  initial  conditions.  The  gray
       area  indicates the  region  of  physical states,  and  the line  the
       concurrence-purity relation for Werner  states. In the inset we observe
       short time deviations from this relation.}
  \label{fig:relCPvarq}
\end{figure}
We thus have an approximate formula for the decay of purity of a Bell pair.
What about concurrence? At this point we take up a result \cite{pineda:012305}
for the behavior of a Bell pair coupled to a kicked spin
chain~\cite{prosenKI}. For a wide range of situations the decay of a pure Bell
state leads to purities and concurrences that closely follow those of a Werner
state in a Concurrence-Purity (CP) diagram~\cite{pineda:012305}. To test model
independence, and thus universality of this behavior we make the corresponding
numerical simulations in the RMT model.  We find that the Werner state CP
relation is quite well fulfilled in the large $N$ limit, as can be seen in
Fig.~\ref{fig:relCPvarq}, where results for fixed coupling but different sizes
of the RMT environment are shown. Studying other couplings allowed by the full
Hamiltonian Eq.~\eqref{eq:hamiltonianboth} leads to similar results even if we
are in different purity decay regimes.  A partial explanation for this
behavior can be found in~\cite{ziman:052325}.

We thus have the second relevant result of this paper; namely the relation of
purity to concurrence for a non-interacting Bell pair interacting with a
chaotic environment follows generically the curve of a Werner state. The
importance of this statement is underlined by the fact, that the actual state
reached at any time is typically not a Werner state.  This is tested, by
considering the spectrum of the density matrix, which should display a triple
degeneracy for a Werner state. In fact a typical spectrum at $P=0.51$ is
$\{0.692,0.142,0.110,0.056\}$.

\begin{figure}
     \includegraphics{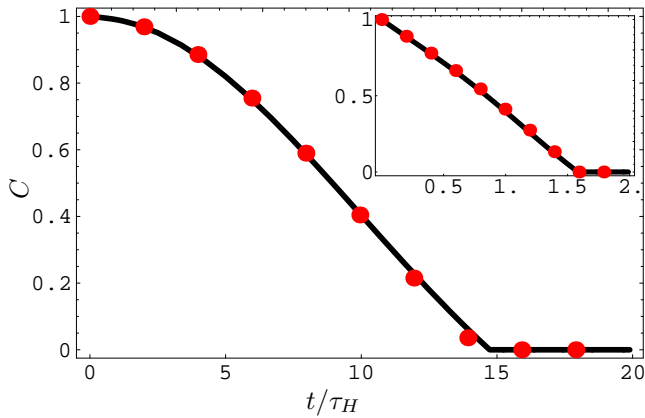}
     \caption{The average  evolution of  concurrence  using and
       environment of dimension 128 and averaging over 10 Hamiltonians and 15
       initial conditions is shown. The black curve corresponds to
       Eq.~\eqref{eq:goodexponwernerCtime}.  We show both the Gaussian regime
       ($\delta=0.008$) and the Fermi golden rule regime in the inset ($\delta=0.07$).}
  \label{fig:longtimeC}
\end{figure}

Having established the genericity of the Werner state relation we can now
insert the expression~\eqref{eq:purityGUE} for purity into the latter and
obtain the heuristic expression
\begin{equation}\label{eq:goodexponwernerCtime}
C_\textrm{ELR}(t)=\max\left\{0, \frac{\sqrt{12P_\textrm{ELR}(t)-3}-1}{2}\right\}
\end{equation}
for concurrence decay.  In Fig.~\ref{fig:longtimeC} we can see that this
relation is well obeyed by Monte Carlo calculations.  Here we obtain
two different time regimes, an exponential one (Fermi golden rule) for strong
perturbations and a Gaussian one for weak perturbation.  The
time scale which defines the crossover between the two regimes is the
Heisenberg time of the environment. Since the exponential behavior can be
obtained letting the Heisenberg time go to infinity, we retrieve results
derived from a master equation approach~\cite{andrereviewshort}.

We have thus obtained a satisfactory expression for concurrence decay, but we
have to remember that we used an extremely simplified model.  In some points
the linear response treatment of the RMT model is slightly affected by these
assumptions; we have made calculations where these approximations where
lifted.  In other words we have allowed both qubits to interact, either with
independent environments or with the same one, see
Eq.~\eqref{eq:hamiltonianboth}.  The form of our result is essentially the
same:
\begin{equation}
P_\textrm{LR}(t)=1-(2-g_\alpha)[\lambda_1^2 f_1(t)+\lambda_2^2 f_2(t)],
\end{equation}
where $f_i(t)$ is identical to $f(t)$ as in Eq.~\eqref{eq:resfinf}, but using
the Heisenberg time of $H_i^\e$.  Other generalizations are possible. Local
Hamiltonians for each qubit causing the degeneracy of the levels of the qubits
to be lifted can be included.  Environment Hamiltonians and couplings chosen
from the Gaussian orthogonal ensemble may also be considered, as well as mixed
states for the environment. Detailed linear response calculations as well as
Monte Carlo calculations for these cases will be presented in the follow up
paper.

Some points are worth mentioning: a) We see significant deviations from the
usual exponential decay at times of the order of the Heisenberg time as
defined by the environment.  Thus, if the spectrum of the environment becomes
very dense, and correspondingly the Heisenberg time moves off to infinity we
recover the usual stochastic result.  b) If the transition region can actually
be seen, then the spectral stiffness of a chaotic environment has a small but
significant stabilizing effect. The absence of spectral stiffness can be
modeled by the so-called Poisson random ensemble~\cite{Dittes}.  c) We
have limited our discussions to the GUE for two reasons.
The simple form of the form factor yields a concise final expression for
purity decay. An additional advantage resulting from the unitary invariance of
the coupling term, is that the final result is invariant under any local
operation at each qubit. This is no longer guaranteed for orthogonal
invariance only. The implications will be discussed in another 
paper.

Summarizing, we have developed a random matrix model for the evolution of a
Bell pair interacting with a generic chaotic environment. Within this model we
derive the linear response approximation for the purity decay of a Bell pair
and show that it differs significantly from decay of a product state of two
spins or qubits, even in the extreme case, where one of the qubits is only a
spectator.  Exponentiation extends the validity of this result far beyond its
original reach.  Monte Carlo calculations show that the relation between
concurrence and purity, as obtained for Werner states, holds for RMT models
and we thus expect it to be generic. Based on these results we have obtained
and  tested   a  heuristic  formula  for   the  decay  of   concurrence  of  a
non-interacting Bell pair.

\begin{acknowledgments}
  We thank T. Gorin, F. Leyvraz, S. Mossmann and T. Prosen for helpful
  discussions. We acknowledge support from projects PAPIIT IN101603 and
  CONACyT 41000F.  C.P. was supported by DGEP.
\end{acknowledgments}

\bibliographystyle{apsrev}
\bibliography{paperdef,miblibliografia,specialbib}

\begin{thebibliography}{20}
\expandafter\ifx\csname natexlab\endcsname\relax\def\natexlab#1{#1}\fi
\expandafter\ifx\csname bibnamefont\endcsname\relax
  \def\bibnamefont#1{#1}\fi
\expandafter\ifx\csname bibfnamefont\endcsname\relax
  \def\bibfnamefont#1{#1}\fi
\expandafter\ifx\csname citenamefont\endcsname\relax
  \def\citenamefont#1{#1}\fi
\expandafter\ifx\csname url\endcsname\relax
  \def\url#1{\texttt{#1}}\fi
\expandafter\ifx\csname urlprefix\endcsname\relax\def\urlprefix{URL }\fi
\providecommand{\bibinfo}[2]{#2}
\providecommand{\eprint}[2][]{\url{#2}}

\bibitem[{spe()}]{specialteleport}
\bibinfo{note}{M. Riebe et al., Nature {\bf 429}, 734 (2004); M. D. Barrett et
  al., Nature {\bf 429}, 737 (2004).}

\bibitem[{\citenamefont{Nielsen and Chuang}(2000)}]{NC00a}
\bibinfo{author}{\bibfnamefont{M.~A.} \bibnamefont{Nielsen}} \bibnamefont{and}
  \bibinfo{author}{\bibfnamefont{I.~L.} \bibnamefont{Chuang}},
  \emph{\bibinfo{title}{Quantum Computation and Quantum Information}}
  (\bibinfo{publisher}{Cambridge University Press},
  \bibinfo{address}{Cambridge, UK}, \bibinfo{year}{2000}).

\bibitem[{\citenamefont{Hill and Wootters}(1997)}]{firstconcurrence}
\bibinfo{author}{\bibfnamefont{S.}~\bibnamefont{Hill}} \bibnamefont{and}
  \bibinfo{author}{\bibfnamefont{W.~K.} \bibnamefont{Wootters}},
  \bibinfo{journal}{Phys. Rev. Lett.} \textbf{\bibinfo{volume}{78}},
  \bibinfo{pages}{5022} (\bibinfo{year}{1997}).

\bibitem[{\citenamefont{Cartan}(1935)}]{cartanRMT}
\bibinfo{author}{\bibfnamefont{{\` E}.}~\bibnamefont{Cartan}},
  \bibinfo{journal}{Abh. Math. Sem. Univ. Hamburg}
  \textbf{\bibinfo{volume}{11}}, \bibinfo{pages}{116} (\bibinfo{year}{1935}).

\bibitem[{\citenamefont{Guhr et~al.}(1998)\citenamefont{Guhr, Mueller-Groeling,
  and Weidenmueller}}]{guhr98random}
\bibinfo{author}{\bibfnamefont{T.}~\bibnamefont{Guhr}},
  \bibinfo{author}{\bibfnamefont{A.}~\bibnamefont{Mueller-Groeling}},
  \bibnamefont{and} \bibinfo{author}{\bibfnamefont{H.~A.}
  \bibnamefont{Weidenmueller}}, \bibinfo{journal}{Phys. Rep.}
  \textbf{\bibinfo{volume}{299}}, \bibinfo{pages}{189} (\bibinfo{year}{1998}),
  \eprint{cond-mat/9707301}.

\bibitem[{\citenamefont{Muller et~al.}(2005)\citenamefont{Muller, Heusler,
  Braun, Haake, and Altland}}]{muller:046207}
\bibinfo{author}{\bibfnamefont{S.}~\bibnamefont{Muller}},
  \bibinfo{author}{\bibfnamefont{S.}~\bibnamefont{Heusler}},
  \bibinfo{author}{\bibfnamefont{P.}~\bibnamefont{Braun}},
  \bibinfo{author}{\bibfnamefont{F.}~\bibnamefont{Haake}}, \bibnamefont{and}
  \bibinfo{author}{\bibfnamefont{A.}~\bibnamefont{Altland}},
  \bibinfo{journal}{Phys. Rev. E} \textbf{\bibinfo{volume}{72}},
  \bibinfo{eid}{046207} (\bibinfo{year}{2005}).

\bibitem[{\citenamefont{Lutz and Weidenmueller}(1999)}]{caoslutz}
\bibinfo{author}{\bibfnamefont{E.}~\bibnamefont{Lutz}} \bibnamefont{and}
  \bibinfo{author}{\bibfnamefont{H.~A.} \bibnamefont{Weidenmueller}},
  \bibinfo{journal}{Physica A} \textbf{\bibinfo{volume}{267}},
  \bibinfo{pages}{354} (\bibinfo{year}{1999}).

\bibitem[{\citenamefont{Gorin and Seligman}(2002)}]{1464-4266-4-4-325}
\bibinfo{author}{\bibfnamefont{T.}~\bibnamefont{Gorin}} \bibnamefont{and}
  \bibinfo{author}{\bibfnamefont{T.~H.} \bibnamefont{Seligman}},
  \bibinfo{journal}{J. Opt. B} \textbf{\bibinfo{volume}{4}},
  \bibinfo{pages}{S386} (\bibinfo{year}{2002}).

\bibitem[{\citenamefont{Gorin et~al.}(2004)\citenamefont{Gorin, Prosen, and
  Seligman}}]{1367-2630-6-1-020}
\bibinfo{author}{\bibfnamefont{T.}~\bibnamefont{Gorin}},
  \bibinfo{author}{\bibfnamefont{T.}~\bibnamefont{Prosen}}, \bibnamefont{and}
  \bibinfo{author}{\bibfnamefont{T.~H.} \bibnamefont{Seligman}},
  \bibinfo{journal}{New J. of Physics} \textbf{\bibinfo{volume}{6}},
  \bibinfo{pages}{20} (\bibinfo{year}{2004}).

\bibitem[{\citenamefont{{R. Sch\"{a}fer et
  al.}}(2005{\natexlab{a}})}]{expRudishort}
\bibinfo{author}{\bibnamefont{{R. Sch\"{a}fer et al.}}},
  \bibinfo{journal}{Phys. Rev. Lett.} \textbf{\bibinfo{volume}{95}},
  \bibinfo{eid}{184102} (\bibinfo{year}{2005}{\natexlab{a}}).

\bibitem[{\citenamefont{{R. Sch\"{a}fer et
  al.}}(2005{\natexlab{b}})}]{njpRudishort}
\bibinfo{author}{\bibnamefont{{R. Sch\"{a}fer et al.}}}, \bibinfo{journal}{New
  J. of Physics} \textbf{\bibinfo{volume}{7}}, \bibinfo{pages}{152}
  (\bibinfo{year}{2005}{\natexlab{b}}).

\bibitem[{\citenamefont{Gorin et~al.}(2006)\citenamefont{Gorin, Seligman, and
  Weaver}}]{gorin-weaver}
\bibinfo{author}{\bibfnamefont{T.}~\bibnamefont{Gorin}},
  \bibinfo{author}{\bibfnamefont{T.~H.} \bibnamefont{Seligman}},
  \bibnamefont{and} \bibinfo{author}{\bibfnamefont{R.~L.}
  \bibnamefont{Weaver}}, \bibinfo{journal}{Phys. Rev. E}
  \textbf{\bibinfo{volume}{73}}, \bibinfo{eid}{015202(R)}
  (\bibinfo{year}{2006}).

\bibitem[{\citenamefont{Zurek}(1991)}]{Zur91}
\bibinfo{author}{\bibfnamefont{W.}~\bibnamefont{Zurek}},
  \bibinfo{journal}{Phys. Today} \textbf{\bibinfo{volume}{44}},
  \bibinfo{pages}{36} (\bibinfo{year}{1991}), \eprint{quant-ph/0306072}.

\bibitem[{\citenamefont{Pineda and Seligman}(2006)}]{pineda:012305}
\bibinfo{author}{\bibfnamefont{C.}~\bibnamefont{Pineda}} \bibnamefont{and}
  \bibinfo{author}{\bibfnamefont{T.~H.} \bibnamefont{Seligman}},
  \bibinfo{journal}{Phys. Rev. A} \textbf{\bibinfo{volume}{73}},
  \bibinfo{eid}{012305} (\bibinfo{year}{2006}).

\bibitem[{bla()}]{blattprivate}
\bibinfo{note}{R. Blatt and H. H\"affner, private communication}.

\bibitem[{\citenamefont{Prosen and Seligman}(2002)}]{purityfidelity}
\bibinfo{author}{\bibfnamefont{T.}~\bibnamefont{Prosen}} \bibnamefont{and}
  \bibinfo{author}{\bibfnamefont{T.~H.} \bibnamefont{Seligman}},
  \bibinfo{journal}{J. Phys. A} \textbf{\bibinfo{volume}{35}},
  \bibinfo{pages}{4707} (\bibinfo{year}{2002}).

\bibitem[{\citenamefont{Prosen}(2002)}]{prosenKI}
\bibinfo{author}{\bibfnamefont{T.}~\bibnamefont{Prosen}},
  \bibinfo{journal}{Phys. Rev. E} \textbf{\bibinfo{volume}{65}},
  \bibinfo{eid}{036208} (\bibinfo{year}{2002}).

\bibitem[{\citenamefont{Ziman and Buzek}(2005)}]{ziman:052325}
\bibinfo{author}{\bibfnamefont{M.}~\bibnamefont{Ziman}} \bibnamefont{and}
  \bibinfo{author}{\bibfnamefont{V.}~\bibnamefont{Buzek}},
  \bibinfo{journal}{Phys. Rev. A} \textbf{\bibinfo{volume}{72}},
  \bibinfo{eid}{052325} (\bibinfo{year}{2005}).

\bibitem[{\citenamefont{{F. Mintert et al.}}(2005)}]{andrereviewshort}
\bibinfo{author}{\bibnamefont{{F. Mintert et al.}}}, \bibinfo{journal}{Phys.
  Rep.} \textbf{\bibinfo{volume}{415}}, \bibinfo{pages}{207}
  (\bibinfo{year}{2005}).

\bibitem[{\citenamefont{{Dittes} et~al.}(1991)\citenamefont{{Dittes}, {Rotter},
  and {Seligman}}}]{Dittes}
\bibinfo{author}{\bibfnamefont{F.-M.} \bibnamefont{{Dittes}}},
  \bibinfo{author}{\bibfnamefont{I.}~\bibnamefont{{Rotter}}}, \bibnamefont{and}
  \bibinfo{author}{\bibfnamefont{T.~H.} \bibnamefont{{Seligman}}},
  \bibinfo{journal}{Phys. Lett. A} \textbf{\bibinfo{volume}{158}},
  \bibinfo{pages}{14} (\bibinfo{year}{1991}).

\end{thebibliography}

\end{document}